\documentclass[12pt,a4paper]{article}
\usepackage{amssymb,amsfonts,amsmath}

\begin{document}

\title{Against identification of contextuality with violation of the Bell inequalities: lessons from theory of randomness}

\author{Andrei Khrennikov\\  Linnaeus University, International Center for Mathematical Modeling\\  in Physics and Cognitive Sciences
 V\"axj\"o, SE-351 95, Sweden}

\maketitle

\abstract{Nowadays contextuality is the hotest topic of quantum foundations and, especially, foundations of  quantum information theory. This notion is characterized by the huge diversity of approaches and interpretations. One of the strongest trends in contextual research is to identify contextuality with violation of the Bell inequalities. We call this sort of contextuality {\it Bell test contextuality} (BTC). In this note, we criticize the BTC-approach. It can be compared with an attempt to identify the complex and theoretically nontrivial notion of randomness with a test for randomness (or a batch of tests, as the NIST-test). We advertize {\it Bohr contextuality} -- taking into account all experimental conditions (context). In the simplest case, the measurement context of an observable $A$ is reduced to joint measurement with a compatible observable $B.$ The latter definition was originally considered by Bell in relation to his inequality. We call it {\it joint measurement contextuality} (JMC). Although  JMC is based on the use of counterfactuals, by considering it in the general Bohr's framework it is possible to handle JMC on physical grounds. We suggest (similarly to randomness) to certify JMC in experimental data with Bell tests, but only certify and not reduce.}

\section{Introduction}

Already Bell pointed out  \cite{BH} that in explanation of violation of Bell type inequalities \cite{BH}--\cite{CHSH}  contextuality is an important alternative to nonlocality. The measurement context  of an observable $A$ was coupled to joint measurement with a compatible observable $B.$ ({\it joint measurement contextuality} - JMC). Thus, by struggling with the nonlocal interpretation of quantum mechanics\footnote{Quantum nonlocality and spooky action at a distance mystify quantum theory. From author's viewpoint 
\cite{NL1,NL3}, quantum nonlocality is only apparent and one can get rid of this misleading notion from quantum theory through the consistent use of Bohr's principle of complementarity \cite{NL0B}.} one can follow Bell and appeal to contextuality \cite{BH}, in the JMC-meaning. 

Unfortunately, in modern quantum foundations, especially in quantum information theory, contextuality  is typically not identified with original Bell's JMC. The theoretically complex and rich notion of contextuality was reduced to one special empirical test, namely violation of the Bell type inequalities ({\it Bell test contextuality} - BTC). In this framework they are known as noncontextual inequalities \cite{CB1}--\cite{CB4}. Of course, one can proceed in this way and be completely fine just by deriving and testing new Bell type (noncontextual) inequalities. However, I am critical to this pragmatic handling of quantum contextuality. It can be compared with an attempt to identify the complex and theoretically nontrivial notion of randomness with one concrete test of randomness (or a few tests) and ignoring deep studies on the notion of randomness, started by von Mises, and continued by Church, Solomonoff,  Kolmogorov, Chatin,  and Martin-L\"ov (see review \cite{QIR} or book \cite{QIRB} for the physicists' friendly presentation). 

Since long time ago (see, e.g. \cite{INT}--\cite{bcont}, I have been advertizing  the notion of contextuality which was considered by Bohr  in his formulation of the 
{\it complementarity principle} \cite{BR}--\cite{PL2}. Bohr did not use the word ``contextuality'' (neither Bell); he wrote about complexes of experimental physical conditions. JMC which role was emphasized by Bell is a special case {\it Bohr contextuality}. And in the Bohr-framework JMC can be handled on the physical (and not metaphysical) grounds. In turn, Bohr contextuality can be rigorously formulated within theory of open quantum systems (see my recent paper \cite{bcont}). Then BTC is just one special class of tests for JMC and generally Bohr contextuality, as say 
NIST-tests for randomness. (This paper is not aimed to diminish the role of experimental tests, neither in theory of randomness nor contextuality.) We shall discuss randomness vs. contextuality tests  in sections \ref{TR}, \ref{TC}.

This is the good place to mention a series of works of Grangier (starting with articles \cite{GRA1}, \cite{GRA2}) in which quantum  contextuality is not reduced to JMC; Grangier's approach is closer to the views of Bohr and the author of this paper. 

We also recall that Bell appealed to contextuality in an attempt to suggest an explanation of violation of his inequality, different from 
nonlocality (see section \ref{BC} for details). Therefore {\it it is meaningless to reduce contextuality to BTC, i.e., to explain violation of the Bell type inequalities by their violation.} 

Quantum contextuality is characterized by diversity of approaches which are not reduced to Bohr, joint measurement, and Bell test  contextualities. However, in this paper we restrict our analysis only to these three approaches. In particular, we do not consider
Kohen-Specker contextuality. (See the recent paper of Svozil \cite{SZR} for detailed review on notions of contextuality.)

\section{Joint measurement contextuality - JMC}

In the discussion on possible seeds of violation of his inequality, Bell argued  \cite{BH} that ``the result of an observation may reasonably depend not only upon the state of the system (including the hidden variables) but also on the complete disposition of the apparatus”. Then Shimony \cite{SH1} emphasized that this is the first statement about contextuality (although Bell did not use this terminology)\footnote{In quantum theory, the word ``contextualistic'' was invented by Shimony \cite{SH2} and a shortening to ``contextual'' was made by Beltrametti and Cassinelli \cite{BEC}. It is surprising that neither Bell nor Shimony even mention Bohr. Did they read Bohr?
(At least Shimony, as philosopher,  should do this...)}.  In fact, Bell’s statement is closely coupled with Bohr’s emphasis of the role of experimental arrangement (see section \ref{BC}). Shimony concreted of the Bell statement  on the role of experimental arrangement  \cite{SH1}:

\medskip

``John Stewart Bell (1928-90) gave a new lease on life to the program of hidden variables by proposing contextuality. In the physical example just considered\footnote{The Bohm version of the EPR experiment - the spin projections measurements}, the complete state $\lambda$ in a contextual hidden variables model would indeed ascribe an antecedent element of physical reality to each squared spin component $s^{2}_n$ but in a complex manner: the outcome of the measurement of $s^{2n}$ is a function $s^{2}_n(\lambda,C)$ of the hidden variable $\lambda$ and the context $C$, which is the set of quantities measured along with $s^{2}_n$. ... a minimum constraint on the context $C$ is that it consists of quantities that are quantum mechanically compatible, which is represented by self-adjoint operators which commute with each other... .''

\medskip

In modern literature the latter sentence is formulated as following.

\medskip

{\bf Definition  (JMC)}.  {\it If $A, B, C$ are three  observables, such that $A$ is compatible with $B$ and  $C,$ a measurement of $A$ might give different result depending upon whether $A$ is measured  with $B$ or with $C.$ }

\medskip
Triple of observable $(A; B, C)$ with given experimental JPDs\\ $p^{\rm{exp}}_{A}, p^{\rm{exp}}_{A}, p^{\rm{exp}}_{B}, p^{\rm{exp}}_{C}$ and $p^{\rm{exp}}_{A B}, p^{\rm{exp}}_{A C}$  is called the JMC scenario.

Using the word ``might'' makes this statement counterfactual. It seems to be difficult (if possible at all) to test JMC experimentally.
Nevertheless, there were published at least two articles which authors claimed that JMC can be tested experimentally and they presented the schemes of experiments \cite{KS3,Griffiths} (which, unfortunately, have never been performed). 

\section{Contextual viewpoint on Bohr's complementarity principle}
\label{BC}

Typically physicists (even the experts in quantum foundations) consider Bohr's writings as difficult for understanding and try to ``simplify'' his statements. In particular, the Bohr's  complementarity principle is widely known as ``wave-particle duality''. First, of we  note that Bohr by himself had never used this notion. In some degree, it is relevant to the earliest attempts of Bohr  (1925) to invent complementarity to quantum physics. However, later Bohr had generalized the notion of complementarity in terms of characteristic properties of quantum measurements. In a long series of publications  \cite{NL0a,NL0B,GG}, I emphasized the contextual 
basis of this ``measurement-complementarity'' principle. It seems that this  (in fact, basic) contextual  component of the complementarity principle was not understood by other experts and quantum contextuality and complementarity go through quantum theory 
relatively independently.  Let cite Bohr (\cite{BR}, v. 2, p. 40-41): 
\medskip

``This crucial point ...  implies {\it the impossibility of any sharp separation between the behaviour of atomic objects and the interaction with the measuring instruments which serve to define the conditions under which the phenomena appear.} In fact, the individuality of the typical quantum effects finds its proper expression in the circumstance that any attempt of subdividing the phenomena will demand a change in the experimental arrangement introducing new possibilities of interaction between objects and measuring instruments which in principle cannot be controlled. Consequently, evidence obtained under different experimental conditions cannot be comprehended within a single picture, but must be regarded as complementary in the sense that only the totality of the phenomena exhausts the possible information about the objects.''
     
The contextual component of this statement can be formulated as the following principle:

 \medskip

{\bf Principle 1 (Contextuality)} {\it  The output of any quantum observable is indivisibly composed of the contributions of the system and the measurement apparatus. Hence, the whole experimental arrangement (context ${\cal C})$ should be taken into account.}

\medskip

Logically, one has no reason to expect that all experimental contexts
can be combined and all observables can be measured jointly. Hence, 
incompatible observables (complementary experimental contexts) can  exist. Moreover, they should exist, otherwise the content of the
contextuality principle would be empty.  Really, if all experimental
contexts can be combined into single context ${\cal C}$ and all observables can be jointly measured in this context, then the outputs of such joint measurements can be assigned directly to a system. To be more careful, we have to say: ``assigned to a system and context ${\cal C}''.$ But, the latter can be omitted, since this is the same context for all observables. 
Thus, {\it contextuality is meaningful only in combination with incompatibility.} 

\medskip

{\bf Principle 2 (Incompatibility)} {\it There exist observables based on complementary experimental contexts. Such observables 
cannot be jointly measured.} 
 
\medskip
Principle 2 is slightly  modified comparing with my previous papers. Observables existing due to Principle 2 are called {\it incompatible.}  Principles 1, 2 can be treated as the integral {\bf Contextuality-Incompatibility  principle.} This is my understanding of the Bohr's complementarity principle. In this paper, we discuss mainly the contextual component of the Contextuality-Incompatibility  principle.

As was mentioned, considered by Bell JMC is the special case of Bohr contextuality. 

\section{Experimental and Kolmogorovian joint probability distributions}

We restrict considerations to observables with finitely many values.

\subsection{Experimental joint probability distribution}
\label{EXP}

Consider a system of physical observables $A, B, C,..., D.$ As we know from quantum physics \cite{BR,VN,PL1,PL2} and cognitive psychology (as well as decision making) \cite{QL1}-\cite{QL4}, some observables can be incompatible, i.e., their joint measurement is impossible 
(see section \ref{BC} for author's reformulation of Bohr's complementarity principle). If some subsystem of observables is jointly measurable then 
experimenters can determine their {\it joint probability distribution} (JPD); for example, let $A$ and $B$ be compatible, then  
$p^{\rm{exp}}_{A B}(a,b)$ can be determined from experiment (as the frequency of  the outcome
$(A=a, B=b)$ in a long  trial of measurements).\footnote{To make this definition mathematically rigorous, one must consider 
infinite sequence of trials (see section \ref{TR}).} It is assumed that,   for each observable, measurement probabilities can be determined:  $p^{\rm{exp}}_{A}(a), p^{\rm{exp}}_{B}(b), ...$ 

Generally there is no consistency between the JPDs of different orders. For example,   even if the observables in each pair 
$A,B$ and $A, C$ are compatible and JPDs $p^{\rm{exp}}_{A B}(a,b), p^{\rm{exp}}_{A C}(a,c)$ are defined, there is no guarantee that the following equalities hold:
\begin{equation}
\label{MAR}
p^{\rm{exp}}_{A}(a) = \sum_b p^{\rm{exp}}_{A B}(a,b) =  \sum_c p^{\rm{exp}}_{A C}(a,c),
\end{equation}
If they hold,  one says that {\it there is no signaling} (for measurement of $A$ jointly with $B$ and with $C$):  measurements of observable $A$ jointly with $B$ and with $C$  generate the same probability distribution  as in measurement of solely $A.$ (The use of the terminology ``signaling'' in such probabilistic formulation might be misleading, but it is widely used quantum physics.) 

We stress that {\it  theoretical probabilities of quantum theory always satisfy to the no signaling condition.} The role of this condition in the EPR-Bohm-Bell experiments was highlighted in the article of Adenier and Khrennikov \cite{AD}. We found that none of the first experiments demonstrating violation of Bell inequalities satisfied to this condition (e.g., \cite{AS0,AS,W}). It was also found that the first experiment which was claimed to be totally loophole free \cite{B3} also suffers of statistically non-negligible signaling \cite{AD1}. Theoretical analysis of this condition was performed in details in articles  of Dzhafarov et al. \cite{DJ}. The classical conditional probabilistic analysis of the condition of no-signaling was done  in paper \cite{AAJ}.

\subsection{Classical probability: observables as random variables}

Consider a system of observables (e.g., physical) $A, B, C,..., D$ with given experimental probability distributions for 
some of its subsystems. The natural questions arises: 
{\it Is it possible to describe them by classical probability theory (CP)?} 
Here ``to described'' means ``in such a way that all experimental probability 
distributions would match the theoretical ones given by CP''.

For readers convenience, we recall the notion of a {\it probability space}. It was invented by Kolmogorov \cite{K} (1933);  it serves as the mathematical basis of the classical probability theory.    Probability space is a triple ${\cal P}=(\Lambda, {\cal F}, p),$ 
where $\Lambda$ is the set of random parameters\footnote{In mathematical literature, typically 
symbol $\Omega$ is used, instead of symbol $\Lambda.$ Points of $\Omega$ are called elementary events. They can be interpreted as realizations of random parameters. In quantum physics random parameters are known as hidden variables.},    ${\cal F}$ is the collection of subsets of $\Lambda$   representing events (this is an event-algebra) and $p$ is the probability measure on   ${\cal F}.$ A {\it random variable} is a function $\xi: \Lambda  \to \mathbf{R}$ such that, for each half-interval  $(a, b],$ the set $\{\lambda\in \Lambda:  a < \xi(\lambda) \leq b\}$ is an event, i.e., it belongs to the set-system ${\cal F}.$
 
In applications of CP to physics,  an observable $A$ is  described by a random variable $A= A(\lambda).$  Consider now compatible (e.g., physical) observables $B_1,...,B_n.$ They can be represented by random variables $B_1= B_1(\lambda),...,B_n=B_n(\lambda).$  Their joint probability distribution (JPD)  is well defined:  
$$
p_{B_1...B_n}(b_1,...,b_n)= p(\lambda \in \Lambda: B_1(\lambda)=b_1,..., B_n(\lambda)=b_n\}.
$$
Any subsystem of these observables is also jointly measurable and the corresponding experimental JPDs approximately equal to the theoretical ones, 
$$
p_{B_1}(b_1) \approx p^{\rm{exp}}_{B_1}(b_1), ...., 
$$
$$
p_{B_1 B_2}(b_1, b_2) \approx p^{\rm{exp}}_{B_1 B_2}(b_1, b_2),....,
$$
and so on. 

Theoretical and, hence, experimental JPDs  for subsystems of observables can be obtained from $p_{B_1...B_n}(b_1,...,b_n)$
as the marginal distribution. For example,
$$
p_{B_1}(b_1)= \sum_{b_2 ... b_n} p_{B_1...B_n}(b_1,...,b_n),....
$$
$$
p_{B_1 B_2}(b_1, b_2)= \sum_{b_3 ... b_n} p_{B_1...B_n}(b_1,...,b_n),....
$$
and so on. We remark that CP-description implies consistency of JPDs of different orders,
for example,
$$
p_{B_1}(b_1)= \sum_{b_2} p_{B_1 B_2}(b_1, b_2) = ...= \sum_{b_2} p_{B_1 B_n}(b_1, b_n).
$$
Hence, there is  no-signaling in the CP-model for observables.

Now turn to the general scheme  of section \ref{EXP}. If physical observables $A, B, C,...,D$  are compatible, then they are CP-representable and have JPD which coincides with the experimental JPD (approximately). Consider now the situation 
such that only some groups of these observables are compatible and for them experimental JPDs can be determined. We are interested in the following question:  {\it Is it possible to represent observables $A, B, C,...,D$ by random variables 
$A=A(\lambda), B=B(\lambda), C=C(\lambda),...,D=D(\lambda)$ on same probability space consistently with experimental JPDs?}  This is the problem of the existence of hidden variables. It is very complicated. Its solution for the special case (CHSH framework) was given by Fine's theorem \cite{Fine, Fine1} (see section \ref{FTEST}). Suppes-Zanotti theorem \cite{SZA} gives the solution for another special case - the original Bell framework for correlated observables \cite{BH, Bell1} (see section \ref{OB}).

\subsection{Existence of triple JPD as noncontextuality test } 

Consider now JMC scenario  $(A; B, C)$ which can realized in the CP-framework, i.e., observables can be represented by random variables
$A=A(\lambda), B= B(\lambda), C= C(\lambda)$ and their JPD $p_{ABC}$ is consistent with the experimental probabilities, 
$p^{\rm{exp}}_{A}, p^{\rm{exp}}_{B}, p^{\rm{exp}}_{C}; p^{\rm{exp}}_{AB}, p^{\rm{exp}}_{AC}$ . 
In the CP-framework, the vector of two random variables $(A,B)$ mathematically  represents the joint measurement of the corresponding observables with the outcome $(a,b),$ where $(a,b)= (A(\lambda), B(\lambda)).$

We point out that the value  $a=A(\lambda)$ depends only on $\lambda$ (``hidden variable''). It does not depend on whether random variable $A$ is considered as a coordinate of the random vector $(A, B)$ or  the random vector $(A, C).$  Hence, the possibility of CP-representation consistent with experimental probabilities is a sufficient condition of noncontextuality.  This framework will be considered (sections \ref{FTEST}, \ref{OB}) for testing JPD existence -- to reject the JMC hypothesis. 

\section{Randomness: notion vs. test}
\label{TR}

The randomness' studies were initiated by von Mises \cite{[169]}--\cite{[171]}.  He  introduced the  notion of a collective which in future 
was formalized as the notion of a random sequence. Let $L = \{\alpha_1,..,\alpha_m\}$ be the set of all possible outcomes of some random experiment (labels in von Mises' terminology); for example, coin tossing with   $L = \{0,1\}.$
A sequence  
\begin{equation}
\label{L5y}
x=(x_1, x_2,..., x_n, ...), x_j \in L,
\end{equation} 
 of experiment's outcomes in a long series of trials is called a collective if it satisfies the following two principles: 
\begin{itemize}
\item statistical stabilization;
\item randomness     
\end{itemize}

 By the first principle for each  $\alpha \in L$ there exists the limit
\begin{equation}
\label{L5t}
p(\alpha; x) =\lim_{N \to \infty} n_N(\alpha)/N,
\end{equation} 
where $n_N(\alpha)$ is the number of $x_j= \alpha$ in the initial block of $x$ of the length $N,$ $(x_1, x_2,..., x_N).$ Per definition this limit is probability of the outcome $\alpha.$ But in this paper we are mainly interested in  the notion of randomness.

Randomness was defined as the limit stability w.r.t. to the special class of selection of subsequences in $x,$ 
so called place selections \cite{[169]}:

{\small ``a subsequence has been derived by a place selection if the decision to retain
or reject the $n$th element of the original sequence depends on the number $n$ and
on label values $x_1,..., x_{n-1}$  of the $n-1$ preceding elements, and not on the
label value of the $n$th element or any following element''.}

Thus a place selection can be defined by a set of functions 
\begin{equation}
\label{psps}
F= \{f_1, f_2(x_1), f_3(x_1, x_2), f_4(x_1, x_2, x_3),..., f_n(x_1,...,x_{n-1}),...\}
\end{equation}
each function yielding the values 0 (rejecting the $n$th
element) or 1 (retaining the $n$th element).
Since any place selection has to produce from an infinite input sequence also an
infinite output sequence, it has also to satisfy the following restriction:
\begin{equation}
\label{pspst}
f_n(x_1,...,x_{n-1})=1 \; \mbox{for infinitely many} \;  n.  
\end{equation}
Here are some examples of place selections: 
\begin{itemize}
\item choose those $x_n$ for which $n$ is
prime; 
\item choose those $x_n$ which follow the word $01;$ 
\item toss a (different) coin;
choose $x_n$ if the $n$th toss yields heads. 
\end{itemize}
Each place selection $F$ is a test of randomness. If a sequence $x$ does not pass some $F$-test, it should be rejected, such a sequence is nonrandom.  To be random, sequence $x$ should pass all place selection tests.     

However, it did not work so simply; one should formalize  the notion of place selection rule more rigorously (see \cite{QIR,QIRB}). This was done by Church  and led to theory of algorithms. However, some Mises-Church random sequences have counter-intuitive properties. The final theory of 
randomness tests was proposed by Martin-L\"of.\footnote{In 1964 and 1965, Martin-Lof studied in Moscow under the supervision of  Kolmogorov. After coming back to Sweden  Martin-L\"of formalized the discussions with his supervisor.} Kolmogorov suggested another approach to randomness based on the notion of algorithmic complexity. This approach was formalized by Chatin.
As was shown by Schnorr, a sequence is Martin-L\"of random if and only it is Kolmogorov-Chatin random. 

Of course, one cannot apply to the concrete sequence $x$ all possible tests for randomness. In applications, people use some batch of tests, e.g., NIST tests.  

Mathematically situation is more interesting. Martin-L\"of showed that there exist the {\it universal test for randomness}. And a sequence is random per definition if it passes this test. However, this proof is not constructive. The result on the existence of  the universal test is similar to the result of Solomonoff-Kolmogorov on the existence of the optimal algorithm used to define the algorithmically random sequence. But, this proof is neither constructive. 
 
\section{Contextuality: notion vs. test}
\label{TC}

Now we turn to JMC. One can find similarity between testing JMC with various Bell-type inequalities and testing randomness
with  various tests for randomness. The crucial difference is that in the latter Solomonoff, Kolmogorov, Martin-L\"of were able to prove the existence of the optimal algorithm and the universal test. This makes theory mathematically rigorous. 
(The impossibility of constructive proofs reminds me the counterfactual nature of JMC.)

\subsection{CHSH-test}
\label{FTEST}

Consider now dichotomous observables taking values $\pm 1.$
and the quadrupole JMC-scenario, given by four triples $(A_i,B_1,B_2), (B_i, A_1, A_2) ,i=1,2.$ Each observable $A_i$ is compatible with both observables $B_i,i=1,2,$ and each observable $B_i$ is compatible with both observables $A_i,i=1,2,$   Thus the their pairwise JPDs are well defined as well as the probability distribution for each observable. {\it No-signaling condition is assumed.}

Now, let us compose the CHSH-correlation     
\begin{equation}
\label{L5r}
{\cal B}_{A_1 A_2; B_1 B_2} = <A_1,B_1> + <A_1, B_2 > +   <A_2, B_1 > - <A_2, B_2>
\end{equation} 
Denote by $\sigma$ some permutation inside $A$ and $B$-blocks or  permutation  of the blocks and consider the corresponding $\sigma$-correlations  ${\cal B}_{\sigma(A_1 A_2; B_1 B_2)}.$ Consider the inequality:
\begin{equation}
\label{L5r}
\max_\sigma |{\cal B}_{\sigma(A_1 A_2; B_1 B_2)}| \leq 2.
\end{equation} 

By Fine's theorem \cite{Fine,Fine1}, there exists the quadrupole JPD 
$
p_{A_1 A_2; B_1 B_2}(a_1, a_2, b_1, b_2) 
$ 
matching the given experimental probabilities iff and only if  inequality (\ref{L5r}) holds true.

Matching of experimental probabilities has the form
$$
p^{\rm{exp}}_{A_1 B_1}(a_1, b_1)= \sum_{a_2, b_2} p_{A_1 A_2; B_1 B_2}(a_1, a_2, b_1, b_2), .... \;
$$
and 
$$
p^{\rm{exp}}_{A_1}(a_1)= \sum_{a_2, b_1, b_2} p_{A_1 A_2; B_1 B_2}(a_1, a_2, b_1, b_2),... 
$$
and so on. 

Existence of this JPD can be considered as mathematical confirmation of noncontextuality.

 The Bell test in the CHSH-form generates the four pairs of sequences of outcomes $\pm 1:$
\begin{equation}
\label{ps}
(x_{A_1j}, y_{B_1j}), \;  (x_{A_1i}, y_{B_2i}), (x_{A_2k}, y_{B_1k}), (x_{A_2m}, y_{B_2m}).
\end{equation}
One puts these four sequences in the expression for the CHSH-correlation  and its permutations and check the condition 
(\ref{L5r}). If it is satisfied, then the experimentally generated quadrupole sequence (\ref{ps}) is rejected, 
it is not contextual (in the JMC-sense). If inequality (\ref{L5r}) is violated, then we say that quadrupole sequence (\ref{ps})
passed  the CHSH-test for JMC contextuality (but nothing more). Thus, the peioneer experiments of Aspect et al. \cite{AS0,AS} and Weihs \cite{W} as well as the `2015-experiments \cite{B3,B1,B2} showed that quadrupole-sequences obtained in them passed the CHSH-test for contextuality.

\subsection{Original Bell's inequality test}
\label{OB}

\subsubsection{Suppes-Zanotti theorem}

Following Suppes and Zanotti \cite{SZA}, onsider three observables $X_1, X_2, X_3$ taking values $\pm 1$ and having zero averages, $E X_j=0.$  Suppose that they are pairwise compatible. Observables in 
each pair $X_1,X_2; X_1,X_3; X_2,X_3$ can be jointly measurable and, hence, their JPDs are well defined,
$p^{\rm{exp}}_{X_1 X_2}, p^{\rm{exp}}_{X_1 X_3}, p^{\rm{exp}}_{X_2 X_3};$ of course, it is assumed that they are separately measurable and their probability distributions are well defined $p^{\rm{exp}}_{X_1}, p^{\rm{exp}}_{X_2}, p^{\rm{exp}}_{X_3}.$ {\it Condition of no-signaling is  assumed.} 

Under the  assumption that  JPD $p_{X_1 X_2 X_3}$ (consistent with the experimental probabilities exists),  Suppes and Zanotti \cite{SZ} derived the following inequality:  
\begin{equation}
\label{SZ}
-1 \leq \langle X_1 X_2 \rangle + \langle X_2 X_3 \rangle + \langle X_1 X_3 \rangle .
\end{equation}
Typically it is identified with the original Bell inequality \cite{}. But, such straightforward interpretation is ambiguous.
If the observables $B_1, B_2,..., B_n$ are quantum and pairwise jointly measurable, then they are jointly measurable and their JPD 
always exists and it is given by Born's rule:
\begin{equation}
\label{L3}
p_{B_1 ... B_n}^{\rm{theor}}(b_1,...,b_n)= \rm{Tr} \hat \rho  \; \hat P_{B_1}^{b_1} ... \hat P_{B_n}^{b_n},
\end{equation}
where $(\hat P_{B_j}^{b_j})$ are the projectors onto the  corresponding eigensubspaces of the operators
 $\hat B_j$ with eigenvalues $ b_j.$  

We will be back to analysis  ``original Bell vs. Suppes-Zanotti  inequalities''.
Now we formulate the Suppes-Zanotti theorem on the existence of triple JPD.

\medskip

{\bf Theorem.} \cite{SZA} {\it  Under  thew above conditions  on observables $X_1, X_2, X_3,$ a  necessary  and  sufficient  condition  for  the  existence  of a triple JPD is that the following  two  inequalities  be  satisfied.} 
\begin{equation}
\label{SZ1}
-1 \leq \langle X_1 X_2 \rangle + \langle X_2 X_3 \rangle + \langle X_1 X_3 \rangle \leq  
 1 + 2 \min\{\langle X_1 X_2 \rangle, \langle X_2 X_3 \rangle, \langle X_1 X_3 \rangle \}
\end{equation}

\subsubsection{From Suppes-Zanotti to original Bell inequality}

The {\it original Bell inequality} has the form \cite{BH,Bell1}: 
\begin{equation}
\label{SZ1}
\langle A_1 B_1 \rangle - \langle A_2  B_2 \rangle + \langle A_1  B_2 \rangle \leq 1.
\end{equation}
Here observable $A_1$ should be compatible with observables $B_1, B_2$ and observable $A_2$ with observable $B_2$
(this is the minimal constraint; typically one assumes that $A_2$ is also compatible with $B_1,$ but the latter is not needed).
 The crucial condition for its derivation is  the precise correlation condition \cite{BH,Bell1}:
\begin{equation}
\label{SZ2a}
<A_2, B_1> = 1 
\end{equation}
(or precise anti-correlation). This inequality differs crucially from other Bell-type inequalities, because of the correlation constraint (\ref{SZ2a}). 

To match with Suppes-Zanotti inequality, we first write (\ref{SZ1}) as
\begin{equation}
\label{SZ1b}
-1 \leq - \langle A_1 B_1 \rangle + \langle A_2  B_2 \rangle - \langle A_1  B_2 \rangle;
\end{equation}
then we change $A_1$ to $-A_1$ and obtain
\begin{equation}
\label{SZ1bt}
-1 \leq \langle A_1 B_1 \rangle +   \langle A_1  B_2 \rangle + \langle A_2  B_2 \rangle.
\end{equation}
Finally, by using the correlation condition (\ref{SZ2a}), we transform (\ref{SZ1b}) into Suppes-Zanotti inequality (\ref{SZ}).
But in an experimental test we should operate with four observables 
$A_i, B_j, i,j, = 1,2.$ So, we reformulate  Suppes-Zanotti's condition (\ref{SZ1})   as
\begin{equation}
\label{SZ1btt}
-1 \leq \langle A_1 B_1 \rangle +  \langle A_1 B_2 \rangle  +\langle A_2 B_2 \rangle\leq  
 1 + 2 \min\{\langle A_1 B_1 \rangle,   \langle A_1 B_2 \rangle, \langle A_2 B_2 \rangle \}
\end{equation}
Thus the triple of observables $A_1, B_1, B_2$ has JPD consistent with experimental probabilities iff inequality (\ref{SZ1b}) holds.
 
The Bell test in the CHSH-form generates the three  pairs of sequences of outcomes $\pm 1:$
\begin{equation}
\label{ps5}
(x_{A_1j}, y_{B_1j}), \;  (x_{A_1i}, y_{B_2i}),  (x_{A_2m}, y_{B_2m}).
\end{equation}
One puts these three sequences in $\langle A_1 B_1 \rangle +  \langle A_1 B_2 \rangle  +\langle A_2 B_2 \rangle$ and check inequality (\ref{SZ1btt}). If it is satisfied, then the experimentally generated triple sequence (\ref{ps5}) is rejected 
as  non contextual (in the JMC-sense). If inequality (\ref{SZ1btt}) is violated, then we say that quadrupole sequence (\ref{ps5})
passed  the CHSH-test for JMC contextuality (but nothing more).

Testing of the original Bell inequality is essentially more complicated than the CHSH-inequality; the additional constraint  
(\ref{SZ2a}) makes the state preparation procedure more complicated. The possibility of such an experimental test was analyzed in author's paper \cite{OBELL} (including analysis of the needed efficiency of detectors). In this paper, the analog of Tsirelson bound for 
the original Bell inequality  (it equals to 3/2) was also found (see also \cite{EL}).  
 .
One can consider a variety of tests based on different Bell type inequalities (see, e.g., \cite{WT}--\cite{A4}.

\section{Concluding remarks}

In this note we criticized identification of BTC with quantum contextuality. In the same way, 
as the notion of randomness is not reduced to a concrete test or a batch of tests, the notion of contextuality 
cannot be reduced to the batch of tests based on the Bell type inequalities. They can only be used to reject the hypothesis on noncontextuality for sequences of outcomes generated by quantum experiment. Although this is important from practical viewpoint, it cannot serve as the basis of the theory of  contextuality. 

The BTC-pragmatism is misleading and this way of thinking slows down development of contextuality theory. {\it The value of BTC for quantum foundations is questionable, BTC explains violation of the Bell inequalities by their violation.}

{\bf Acknowledgments} 
This paper is a part of the special issue devoted to memory of Vladimir Andreev. We had many exciting discussions during 
Vladimir's visits to V\"axj\"o (including discussions on paper \cite{A1}--\cite{A4}). Although our views on the possible interpretations of the violation of the Bell inequalities 
did not coincide, such discussions were extremely useful for structuring my views on violations of these inequalities.

The paper was written 
with the partial financial support of the Ministry of Education and Science of
the Russian Federation as part of the program of the Mathematical Center
for Fundamental and Applied Mathematics under the agreement
N 075-15-2019-1621 .

\end{document}